\pdfoutput=1
\documentclass[runningheads]{llncs}

\usepackage[T1]{fontenc}
\usepackage{lmodern}
\usepackage{microtype}
\usepackage{graphicx}
\usepackage{booktabs}
\usepackage{amsmath}

\usepackage{tikz}
\usetikzlibrary{positioning, arrows.meta, shapes.geometric, fit,
                calc, backgrounds, decorations.pathreplacing}

\pdfpagewidth=\paperwidth
\pdfpageheight=\paperheight

\usepackage{hyperref}
\usepackage[capitalize]{cleveref}
\crefname{section}{Sect.}{Sects.}
\Crefname{section}{Section}{Sections}
\crefname{figure}{Fig.}{Figs.}
\Crefname{figure}{Figure}{Figures}
\crefname{table}{Table}{Tables}
\Crefname{table}{Table}{Tables}

\usepackage{eso-pic}

\newcommand\copyrighttext{%
\footnotesize
This manuscript has been accepted for presentation at the
2nd International Conference on Agentic and Generative Techniques
in Intelligent Computational Systems (AGENTICS) 2026,
held in Angers, France, 28 -- 30 October 2026, and for publication
in Springer Communications in Computer and Information Science (CCIS) proceedings.
This is the author's accepted manuscript version.
The final authenticated publication will be available via Springer.
}
\newcommand{\copyrightnotice}{%
\AddToShipoutPictureFG*{%
\AtPageLowerLeft{%
\raisebox{1cm}{%
\makebox[\paperwidth][c]{%
\fbox{\parbox[b]{0.9\textwidth}{\copyrighttext}}%
}}}}}

\graphicspath{{figures/}}

\begin{document}

\copyrightnotice

\title{The Agentic Company OS: Substrate Inversion for Sustained
  Enterprise Agent Deployment}
\titlerunning{The Agentic Company OS}
\author{Oliver Aleksander Larsen\inst{1}\orcidID{0009-0003-6944-2560}$^{\star}$\\
Mahyar T. Moghaddam\inst{1}\orcidID{0000-0001-5028-7546}}
\authorrunning{O. A. Larsen and M. T. Moghaddam}
\institute{SDU Software Engineering, University of Southern Denmark,
  Odense, Denmark\\
  \email{olar@mmmi.sdu.dk$^{\star}$, mtmo@mmmi.sdu.dk}}
\maketitle

\begin{abstract}
Enterprise AI agents often succeed in a demonstration and then
stall once they must operate day after day. An industry report
estimates that most pilots never reach production and that
deployed systems rarely retain feedback or improve over time,
while agent benchmarks show single-run successes masking
unreliable repetition. We argue that these failure modes share a
common architectural root: agents reason over data structured for
human operators and traditional applications, not for the language
models that power them. This position paper proposes that
companies deploying agents in sustained operation should rebuild
their cognitive substrate, the shared environment agents
read as working context, around representations matched to that
reasoning surface, isolating schema translation to the action
boundary. Markdown is the instantiation available today, not a
proven agent-native primitive. Two mechanisms ground the
argument: context-bandwidth asymmetry, the gap between one-pass
reading of connected prose and field-by-field typed access that
strips relations; and cross-loop coupling, the claim that action,
skill, and policy loops compound only if they share one
substrate. A four-layer framework (Data, Knowledge, Intelligence,
Governance) operationalizes the position, with a Sync Agent
enforcing the action boundary and a per-skill trust gradient,
making governance and auditability structural properties of the
substrate. The position revives the shared-substrate tradition of
classical multi-agent systems under LLM-era economics. We analyze
the main objections and risks, including indirect prompt
injection on the compile path, and outline a research agenda for
evaluating substrates directly.
\keywords{LLM agents \and Agentic AI \and Agent governance \and
  Enterprise architecture \and Business process management \and
  Prompt injection \and Multi-agent coordination}
\end{abstract}

\section{Introduction}
\label{sec:intro}

Enterprise AI adoption is scaling rapidly \cite{aiindex2025}, and
agentic systems are entering production deployment
\cite{staufer2026agentindex}. Yet a widely circulated industry
report estimates that only a small fraction of enterprise GenAI
pilots reach production \cite{challapally2025genai},
and the single successful run of a demonstration masks
inconsistency that repeated operation exposes \cite{yao2025taubench}.
We identify five recurring failure modes, often treated in
isolation: governance gaps, where an agent acts outside policy;
transparency gaps, where reviewers cannot reconstruct agent actions;
coordination gaps, where agents drift out of step; safety gaps,
where untrusted content contaminates trusted reasoning; and
plateaued improvement, where deployments stop improving despite
model upgrades. An index of deployed agents documents how sparsely
governance, transparency, and safety practices are disclosed
\cite{staufer2026agentindex}. The industry report also finds that
most deployed GenAI systems fail to
retain feedback or improve over time \cite{challapally2025genai},
and model version changes can even degrade behavior
\cite{chen2024chatgpt}.
The motivating hypothesis is that this stall is architectural: it
hardens into a plateau unless the agent's working context is rebuilt.

We propose that these are not five orthogonal problems but five
symptoms with a common architectural root: agents reason over a
substrate, the representation an agent reads as working context,
built for human operators and conventional applications, not for
the language models doing the reasoning. Rebuilding the substrate
around the agent's reasoning surface would address four of the
five directly (governance, transparency, coordination, and
plateaued improvement); the safety gap it reshapes rather than
removes (\cref{sec:safety}).

\textbf{We argue that companies deploying AI agents in sustained
operation should rebuild their cognitive substrate (the shared
environment the agent reads as working context) around
representations matched to that reasoning surface, with schema
translation isolated to the action boundary.} The position is a
structural commitment about where agent reasoning happens, not a
framework, runtime, or model-class recommendation. Markdown is the
instantiation available today, not a proven agent-native primitive
(matched to how an LLM reads connected prose rather than typed
application schemas).
Any representation better matched to an LLM's reasoning surface
would serve equally well. We do not claim this architecture has
been shown to outperform typed-tool, retrieval, or hybrid stacks;
\cref{sec:call} specifies the comparison that would. We call this
commitment substrate inversion. It renews the shared-substrate
tradition of classical multi-agent systems under LLM-era economics,
a lineage that \cref{sec:background} develops.

Three trends make the argument timely. First, one million tokens of
LLM context \cite{comanici2025gemini25} is on the order of two
thousand pages of Markdown, enough to hold the compiled knowledge
base of a small or mid-sized company. Second, closed-loop measurement
is becoming cheap: a thousand outcome records fit in a single
context window, and an LLM can draft a playbook revision from them
in one pass. Third,
deployments are moving past chatbot scope into autonomous operation,
so substrate choices can no longer be deferred.

The paper makes three contributions, with relevance to business
process management, contextual decision-making, and applicative
risk identification.
First, it states the substrate-inversion position, grounded in
two mechanisms: context-bandwidth asymmetry (one-pass connected
prose versus field-by-field typed access that strips relations)
and cross-loop coupling (action, skill, and policy loops compound
only if they share one substrate).
Second, it presents a four-layer framework (Data, Knowledge,
Intelligence, Governance) operationalizing the position, with a Sync
Agent at the action boundary, a per-skill trust gradient, and three
closed improvement loops sharing one substrate. Third, it critically
examines the strongest objections (\cref{sec:alternatives}) and
risks, including indirect prompt injection (\cref{sec:safety}),
deriving an agenda for evaluating substrates rather than agents,
with falsifiable hypotheses and an experiment design to test them
(\cref{sec:call}). The applicable domain is
enterprise operations where interpretation-heavy work dominates.
The framework distills an ongoing design effort for an agent-native
company operating system, presented as an arguable architectural
thesis open to empirical test.

\Cref{sec:background} situates the position in classical
multi-agent systems and in three families of current enterprise
agent stacks. \Cref{sec:position} states the two mechanisms and
the hypotheses they yield. \Cref{sec:framework} presents the
four-layer framework and walks through a constructed one-page
customer file that both a human and an agent read as shared
context (\cref{sec:example}). \Cref{sec:alternatives} takes three
alternative approaches at their strongest;
\cref{sec:implications} names imported risks and scope
conditions; \cref{sec:call} gives an adoption path and a
substrate-controlled benchmark.

\section{Background and Related Work}
\label{sec:background}

Software agents that perceive, reason, and act have been studied
since the 1980s, from blackboard systems such as Hearsay-II
\cite{erman1980hearsay} through the mid-1990s surveys that named
the field \cite{wooldridge1995agents}. What changed with large
language models is not the idea of an agent but the form of its
working context: a natural-language window rather than a
hand-engineered memory. Enterprise agent stacks built on that
surface fall into three families. Retrieval-augmented agents over
relational systems \cite{fan2024ragsurvey,lewis2020rag} retrieve
fragments from CRMs, document stores, and knowledge bases and
prompt the model with them. The working context still comes from
the relational store.
Tool-using agents on typed schemas compose function calls into
multi-step workflows through ReAct-style loops \cite{yao2023react}.
Here the
working context still comes from the schema-locked system behind
the typed contracts (fields and relations fixed by the
application's type system). Vertical AI
applications ship domain-specific copilots (support, sales, legal,
finance), each with a store optimized for its vertical. All
three families deliver value on bounded tasks, and benchmark
evidence is consistent with a ceiling tied to the working
context, not only to the model: GAIA
performance drops with task depth \cite{mialon2024gaia}, and
AgentBench \cite{liu2024agentbench} reports persistent multi-step
gaps, both confounded with planning and tool limits. A recent
file-native evaluation finds serialization format alone does not
significantly affect single-task accuracy
\cite{mcmillan2026filenative}; what no benchmark yet isolates is
substrate structure (connected prose versus fragmented access) or
improvement that compounds over repeated operation, which
motivates the evaluation proposed in
\cref{sec:call}. The first two families inherit the
working context from the human-first system the agent was attached
to, and vertical applications choose theirs only per vertical.
Translation compensates for
the resulting mismatch but does not remove it (\cref{tab:compare}).
We reject that inheritance: the working context should be rebuilt
for the agent, not borrowed from the human-first system and
patched with retrieval or tools.

\begin{table}[t]
\centering
\caption{Where an agent reasons, where schema translation is paid,
whether improvement loops share a substrate, and where audit and
governance live. The three deployed families inherit a human-first
store and keep policy and provenance outside it; substrate
inversion rebuilds the working context and places translation,
audit, and policy on one versioned corpus (\cref{sec:position}).}
\label{tab:compare}
\footnotesize
\setlength{\tabcolsep}{4pt}
\begin{tabular}{@{}lp{2.05cm}p{2.05cm}p{2.05cm}p{2.05cm}@{}}
\toprule
 & \raggedright RAG agents & \raggedright Typed-tool agents
  & \raggedright Vertical apps & \raggedright Substrate inversion
  \tabularnewline
\midrule
Reasoning surface & \raggedright retrieved fragments
  & \raggedright typed results & \raggedright vertical schema
  & \raggedright compiled corpus \tabularnewline
Schema translation & \raggedright retrieval pipeline
  & \raggedright every call & \raggedright app adapters
  & \raggedright action boundary \tabularnewline
\begin{tabular}[t]{@{}l@{}}Loops on one\\ substrate\end{tabular}
  & none & action only & per vertical & all three \tabularnewline
Audit-trail locus & source logs & API logs & vendor logs
  & \raggedright version control \tabularnewline
Governance locus & \raggedright external engine
  & \raggedright external engine & \raggedright vendor controls
  & \raggedright substrate itself \tabularnewline
\bottomrule
\end{tabular}
\end{table}

Three theoretical anchors support the position. Sculley et al.
\cite{sculley2015} describe hidden technical debt in machine
learning systems as glue code, pipeline jungles, and feedback loops
that age badly. Translation layers between agents and operational
ground truth are the agent-era form of that debt.
Conway's
observation that organizations produce systems mirroring their own
communication structure \cite{conway1968} has a direct agent-era
reading: agent
organizations inherit the schema fragmentation of the org
chart. Hutchins's analysis of
distributed cognition \cite{hutchins1995} argues that
multi-actor cognition runs through shared external representations.
The position extends this claim from coordination to closed-loop
improvement, because corrections travel along the representations
that already coordinate human and agent collaborators.

The architecture family itself predates the LLM. Blackboard systems
coordinated specialist problem solvers through a shared working
memory \cite{erman1980hearsay,nii1986blackboard}. Classical MAS
explored shared media and institutions (Linda, environment-mediated
coordination, electronic institutions, adjustable autonomy, and
mediators
\cite{gelernter1985linda,weyns2007environment,esteva2001institutions,scerri2002adjustable,wiederhold1992mediators})
over two decades. These designs were structurally right and economically
blocked: every shared representation had to be hand-engineered into
frames, tuples, or ontologies before any agent could read it. The
other classical blocker was the blackboard control problem, deciding
which knowledge source should act next \cite{nii1986blackboard}. The
LLM removes the first blocker and relocates the second: plain prose
is now machine-readable at the scale of a company, while scheduling
falls to an orchestrator agent whose control policy remains open
(\cref{sec:conclusion}). That reversal makes the blackboard pattern
economically plausible as enterprise infrastructure.

Several recent artifacts inform the framework. Karpathy's
living-wiki proposal \cite{karpathy2026} gives the substrate idea
its current instantiation: raw data feeds an LLM compiler that
emits and maintains a wiki of Markdown files, which the model
navigates index-first, reading pages as needed instead of
retrieving fragments at query time.
LLM-based multi-agent systems, surveyed by Guo et al.
\cite{guo2024llmmas}, provide scaffolding for early prototypes. Agent memory systems
\cite{packer2023memgpt} solve the persistence problem
inside the runtime; substrate inversion solves it in the shared
substrate.
Practitioners in business process management report demand for
governance guardrails on agents \cite{vu2025agenticbpm}.

The position engages the agent-safety and prompt-injection
literature in two directions. In the first, it contributes safety
infrastructure: the audit trail is the version-control log, the
trust gradient is recorded in versioned skill metadata, and policy
lives in the same Markdown the agents read. In the
second, it imports a genuine threat surface: compiling untrusted
external content into trusted agent context exposes the
architecture to prompt-injection attacks
\cite{greshake2023,perez2022}, with indirect injection as the most
serious case. \Cref{sec:safety} examines this risk.

Recent work treats one piece of the classical assembly at a time:
a governance regime \cite{shavit2023governing},
a learning method \cite{wang2023voyager}, a safety control
\cite{hines2024spotlighting}, and blackboard coordination itself,
revived as a runtime pattern for LLM multi-agent systems
\cite{han2025blackboard,salemi2025blackboard}. To our knowledge,
no prior or concurrent work assembles the whole: an enterprise
prose substrate, governance and audit in version control, and
improvement loops coupled through one corpus.

\section{Substrate Inversion: The Argument and Its Mechanisms}
\label{sec:position}

The position commits to two structural properties: a single
agent-native reasoning surface, and schema translation isolated to
the action boundary. The two mechanisms below establish why these
properties, not the implementation details around them, are the
load-bearing choices. The multi-agent
systems community has long treated the environment as a first-class
abstraction \cite{weyns2007environment}. Substrate inversion
applies that argument to LLM agents: the substrate is the
environment, and choosing it for the agent, rather than inheriting
it from the legacy application, is the central design decision.

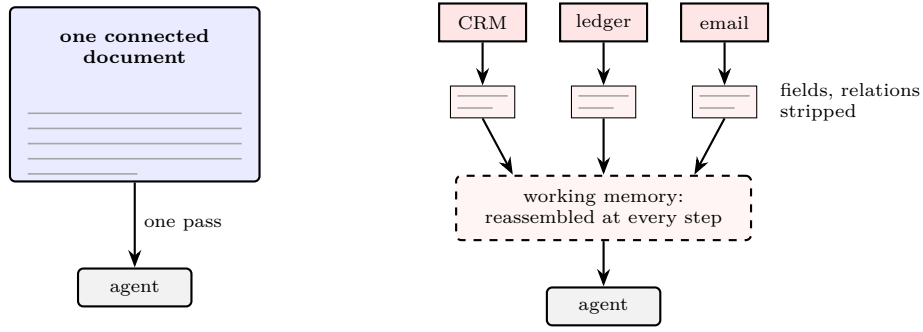
\begin{figure}[t]
\centering
\begin{tikzpicture}[
  doc/.style={rectangle, draw, thick, fill=blue!8, rounded corners=2pt,
    minimum width=3.3cm, minimum height=2.3cm},
  sys/.style={rectangle, draw, thick, fill=red!10,
    minimum width=1.15cm, minimum height=0.5cm, font=\scriptsize},
  frag/.style={rectangle, draw, thin, fill=red!5,
    minimum width=0.85cm, minimum height=0.42cm},
  wm/.style={rectangle, draw, thick, dashed, fill=red!4,
    rounded corners=2pt, minimum width=3.9cm, minimum height=0.85cm,
    align=center, font=\scriptsize},
  ag/.style={rectangle, draw, thick, fill=black!5, rounded corners=2pt,
    minimum width=1.5cm, minimum height=0.5cm, font=\scriptsize},
  flow/.style={->, thick, >=Stealth, font=\scriptsize},
  tline/.style={black!35, line width=0.6pt}
]
\node[doc] (docnode) at (0,0.2) {};
\node[align=center, font=\scriptsize, anchor=north]
  at ($(docnode.north)+(0,-0.18)$)
  {\textbf{one connected}\\\textbf{document}};
\foreach \y in {-0.05,-0.25,-0.45,-0.65}
  \draw[tline] ($(docnode.west)+(0.25,\y-0.2)$)
    to ($(docnode.east)+(-0.25,\y-0.2)$);
\draw[tline] ($(docnode.west)+(0.25,-1.05)$)
  to ($(docnode.west)+(1.7,-1.05)$);
\node[ag] (agentl) at (0,-2.35) {agent};
\draw[flow] (docnode) to node[right] {one pass} (agentl);

\node[sys] (crm) at (4.6,1.15)  {CRM};
\node[sys] (led) at (6.2,1.15)  {ledger};
\node[sys] (eml) at (7.8,1.15)  {email};
\node[frag] (fa) at (4.6,0.1) {};
\node[frag] (fb) at (6.2,0.1) {};
\node[frag] (fc) at (7.8,0.1) {};
\foreach \f in {fa,fb,fc} {
  \draw[tline] ($(\f.west)+(0.1,0.08)$)  to ($(\f.east)+(-0.1,0.08)$);
  \draw[tline] ($(\f.west)+(0.1,-0.08)$) to ($(\f.east)+(-0.3,-0.08)$);
}
\draw[flow] (crm) to (fa);
\draw[flow] (led) to (fb);
\draw[flow] (eml) to (fc);
\node[font=\scriptsize, align=left, anchor=west] at (8.42,0.1)
  {fields, relations\\stripped};
\node[wm] (reasm) at (6.2,-1.3)
  {working memory:\\reassembled at every step};
\draw[flow] (fa.south) to ($(reasm.north)+(-1.2,0)$);
\draw[flow] (fb.south) to (reasm.north);
\draw[flow] (fc.south) to ($(reasm.north)+(1.2,0)$);
\node[ag] (agentr) at (6.2,-2.6) {agent};
\draw[flow] (reasm) to (agentr);
\end{tikzpicture}
\caption{Context-bandwidth asymmetry in plain terms: on the left,
the agent reads one connected document in a single pass, with
relations carried in the prose; on the right, typed systems return
fields that strip those relations, so the agent rebuilds the
narrative in working memory at every step. Hypothesis H1 predicts
that this reassembly cost bounds compositional reach.}
\label{fig:reassembly}
\end{figure}

\textbf{Mechanism 1: context-bandwidth asymmetry.} Modern LLM
context windows accept roughly $10^{6}$ tokens of unstructured
prose \cite{comanici2025gemini25}, consumed in a single pass with
relational context carried implicitly. Typed APIs and relational
schemas expose the same information field by field, with relations
declared in the schema rather than carried in the content
(\cref{fig:reassembly}). Each schema crossing therefore strips
relational context that the agent must rebuild in working
memory. Consider a multi-step sales decision
that depends on five contexts: the prospect's most recent reply,
the deal's pipeline stage, the customer's payment history, the
prospecting playbook in force, and the company's pricing rules. On
a Markdown substrate these arrive as a connected narrative
read in one pass. Routed through a CRM, a ledger, and an email API,
they arrive as typed records, and the agent reassembles the
narrative anew at each step.

The loss is depth-dependent: the deeper the compositional task, the
more the per-step reassembly compounds. Even information present in
context is recovered unreliably depending on its position, a limit
one-pass reading shares \cite{liu2024lost}.
We state the
resulting prediction as hypothesis H1: per-step reassembly cost
bounds an agent's compositional reach, a bound that tightens with
task depth. The benchmark in \cref{sec:call} tests it.

\textbf{Mechanism 2: cross-loop coupling.} We model sustained
improvement as three feedback loops at three time-scales: an action loop
on seconds (act, observe, adjust), a skill loop on days
(accumulated outcomes revise the playbook), and a policy loop on
weeks or months (humans or governance agents detect systemic
patterns and change the rules).
The loops are coupled: a corrected outcome compounds only if it
propagates from action to skill to policy, and a policy update
takes effect only when the next action reads it. The loop
structure echoes autonomic computing's MAPE-K and the three-layer
architecture of self-managed systems
\cite{kephart2003autonomic,kramer2007selfmanaged}; substrate
inversion externalizes the knowledge component as a prose corpus
that humans, agents, and governance share. Enterprise stacks often
split them across three substrates: the action
loop in agent memory, the skill loop in fine-tuning artifacts or
vendor prompt registries, and the policy loop in meeting notes.
A study of teams building product copilots documents fragmentation
of this kind \cite{parnin2025copilot}, echoing the technical debt long
described for classical ML \cite{sculley2015}. A
correction noticed at action time then fails to migrate to skill
time, a policy decision fails to alter the next call, the loops
desynchronize, and corrections stop compounding.

Linda-style coordination models decouple writers and readers in
time and identity \cite{gelernter1985linda}; cross-loop
coupling exploits the same decoupling, since a policy written
at week scale is read at action scale. A single Markdown substrate is the kind of shared
external representation that distributed cognition theory
\cite{hutchins1995} describes: the playbook the agent reads next
is the policy committed in the last governance review. This yields hypothesis H2,
the paper's central falsifiable claim: under split substrates,
improvement decays toward the noise floor rather than compounding
as it does on a unified substrate, and model upgrades alone do not
rescue it. The same benchmark in \cref{sec:call} tests it.

Together the two mechanisms point to the full substrate-inversion
commitment: one representation matched to the LLM consumer, shared
across the three improvement loops, with translation paid once at
the periphery.

\section{The Agentic Company OS Framework}
\label{sec:framework}

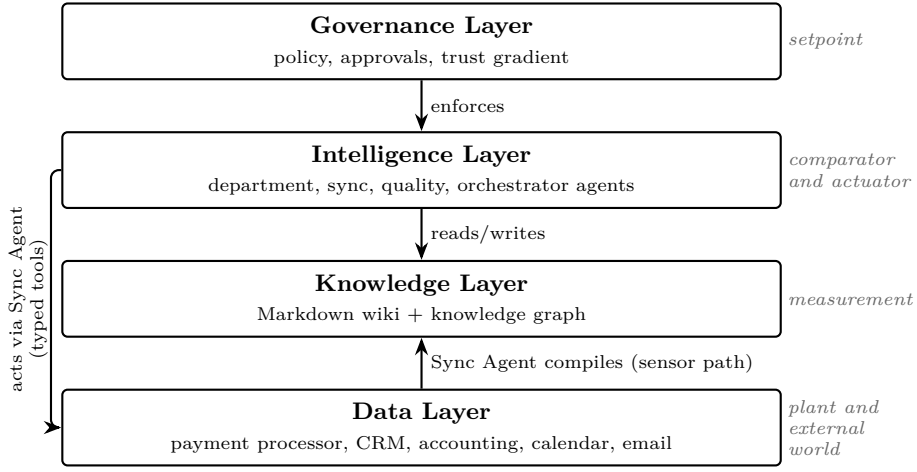
\begin{figure}[t]
\centering
\begin{tikzpicture}[
  layer/.style={
    rectangle, draw, thick, rounded corners=2pt,
    minimum width=9.5cm, minimum height=1.0cm,
    align=center, font=\small
  },
  flow/.style={->, thick, >=Stealth, font=\scriptsize},
  role/.style={font=\scriptsize\itshape, text=black!60, anchor=west,
    align=left, inner sep=1pt}
]
\node[layer] (gov)   at (0, 5.1) {\textbf{Governance Layer}\\
  \scriptsize policy, approvals, trust gradient};
\node[layer] (intel) at (0, 3.4) {\textbf{Intelligence Layer}\\
  \scriptsize department, sync, quality, orchestrator agents};
\node[layer] (know)  at (0, 1.7) {\textbf{Knowledge Layer}\\
  \scriptsize Markdown wiki + knowledge graph};
\node[layer] (data)  at (0, 0.0) {\textbf{Data Layer}\\
  \scriptsize payment processor, CRM, accounting, calendar, email};

\node[role] at (4.80, 5.1) {setpoint};
\node[role] at (4.80, 3.4) {comparator\\and actuator};
\node[role] at (4.80, 1.7) {measurement};
\node[role] at (4.80, 0.0) {plant and\\external\\world};

\draw[flow] (gov) to node[right] {enforces} (intel);
\draw[flow] (intel) to node[right] {reads/writes} (know);
\draw[flow] (data) to node[right] {Sync Agent compiles (sensor path)}
  (know);

\coordinate (side) at (-4.90, 1.7);
\draw[flow, rounded corners=3pt] (intel.west) -| (side) |- (data.west);
\node[rotate=90, anchor=south, align=center, font=\scriptsize,
  inner sep=1pt] at (side) {acts via Sync Agent\\(typed tools)};
\end{tikzpicture}
\caption{The four-layer architecture as a closed control loop.
Governance holds the setpoints, Intelligence compares and acts,
Knowledge is the measurement substrate, and the Sync Agent is the
only translation at the action boundary (sensor and actuator).
Action, skill, and policy loops compound only because they read
and write the same Knowledge Layer (cross-loop coupling).}
\label{fig:architecture}
\end{figure}

The four layers exist because each has a job the others cannot
absorb. The Data Layer stays the company's authoritative systems
of record: transactional integrity and statutory record-keeping
are orthogonal to reasoning, and demanding both from one store is
what produces schema fragmentation. The Knowledge Layer is the
reasoning surface: compiled, connected prose that humans and
agents read as working context. The Intelligence Layer
coordinates by writing those shared files, so there is no
out-of-band message bus. The Governance Layer lives in the same
files, so policy cannot drift from operational state. The Sync
Agent exists for one reason: it is the only translation at the
action boundary, compiling into the Knowledge Layer and executing
typed writes back out. A trust gradient (\cref{sec:trust}, \cref{fig:trust}) makes
governance structural, and three closed loops (\cref{sec:loops})
compound through the shared substrate
(\cref{sec:layers}, \cref{fig:architecture}).

Open operational limits, including freshness of compiled files,
read-level authorization, write concurrency, cost and latency of
rereading, and recovery from substrate corruption, are design
caveats rather than solved claims; \cref{sec:call} lists them as
questions a deployment must answer.

\subsection{The Four Layers}
\label{sec:layers}

\textbf{Data Layer.} The Data Layer is the infrastructure the
company already runs: the payment processor,
the CRM, the accounting system, the calendar, the email server.
These systems are schema-locked. Agents visit them only through
the Sync Agent, when an action needs the guarantees of statutory
record-keeping or an investigation needs the raw records. The framework leaves these systems
authoritative and unmodified, because their job (transactional
integrity, regulatory compliance) is orthogonal to the agent's
job (reasoning), and demanding both from one substrate is what
produces schema fragmentation. The integration
glue (webhooks, credentials, write-backs) relocates into the Sync
Agent rather than disappearing.

\textbf{Knowledge Layer.} The Knowledge Layer is a single Git
repository of Markdown files, organized as a wiki with
bidirectional links and an LLM-compiled knowledge graph
\cite{edge2024graphrag}, in the spirit of recent agent-native
proposals \cite{karpathy2026}. Decisions, summaries, playbooks,
and contextual narrative live here; raw transactional rows do not.
Humans and agents read the same files and leave their traces in
the same Git history. In agents-and-artifacts terms
\cite{omicini2008artifacts}, customer files, playbooks, and policy
documents are coordination artifacts: first-class entities that
mediate agent work and persist beyond any single interaction. This
is the reasoning surface hypothesized by Mechanism 1, not a
measured gain over retrieval, and it stays small: at one page per customer and per
playbook, a 50-person company with a few hundred accounts compiles
to a few hundred thousand tokens, inside one context window.
Beyond that envelope, at thousands of accounts, one-pass reading
gives way to index-first navigation over the same substrate, and
the bandwidth argument applies per page set: each page is a
connected narrative, not a stripped fragment.

\textbf{Intelligence Layer.} The Intelligence Layer is a small set
of specialized agents (department, sync, quality, orchestrator)
that share the Knowledge Layer as working memory. Agents exchange
no out-of-band messages; they coordinate by writing to shared
files, and the substrate serves as the message bus. This revives the blackboard and tuple-space
patterns \cite{gelernter1985linda,nii1986blackboard}, with one
decisive change: the shared workspace is prose that the LLM and
the human read natively, not an engineered representation that
only purpose-built knowledge sources could parse. Write concurrency
follows a minimal regime: the Sync Agent alone writes compiled
frontmatter (the YAML header it compiles), the skill loop revises
playbook bodies while trust fields accept no LLM-agent writes
(a commit-gate rule; \cref{sec:trust}), agents append to activity
logs rather than
editing prose in place, and the orchestrator serializes residual
same-file edits.

\textbf{Governance Layer.} The Governance Layer is the policy
plane: rules, approval thresholds, and the per-skill trust
gradient (\cref{sec:trust}), all written in the same Markdown the
agents read. Every commit is signed and revertable, and every
policy change is itself an artifact in the audit trail. No
separate policy engine can drift from operational state, because
the policy engine is the operational state, echoing
law-governed interaction \cite{minsky2000lgi} and the
normative rules of electronic institutions
\cite{esteva2001institutions}. Legibility is in-band by
construction; enforcement sits at the action boundary
(\cref{sec:safety}).

\subsection{The Sync Agent and the Action Boundary}
\label{sec:sync}

The Sync Agent is the only agent permitted to call external
systems, playing the classical mediator role
\cite{wiederhold1992mediators} at the action boundary. Two
ingestion modes feed the Knowledge Layer. \emph{Continuous
ingestion} combines webhooks for structured events with a single forwarding
email address that catches the unstructured long tail (meeting
notes, ad-hoc updates).
\emph{Batch ingestion} is a one-time
migration at onboarding: a script pulls historical data through
existing read-only API keys and writes summary files into the
right folders. Continuous ingestion keeps the Knowledge Layer fresh
without polling, and quality agents reconcile compiled files against
the systems of record via the Sync Agent, flagging missed events and hallucinated
summaries for recompilation: compiled state is a materialized
view, never authoritative.

The Sync Agent writes summaries in the Knowledge Layer's
representation, not raw rows in the source schema. A customer with
dozens of invoice lines and
activity entries in the CRM appears in the substrate as a single
one-page file. The full rows remain in the systems of record, where law
and audit need them. An agent that requires a deep investigation
calls typed tools via the Sync Agent and writes a fresh
summary back afterwards.
The Model Context Protocol, surveyed in
\cite{hou2025mcp}, standardizes typed tool contracts at this
boundary, at the periphery, not in the reasoning loop.
Recompiling a few-hundred-thousand-token Knowledge Layer daily
costs single-digit dollars on a mid-tier long-context model;
continuous reasoning likely dominates the budget, since rereading
the full corpus on every action multiplies that cost and adds
prefill latency to the seconds-scale action loop. Prefix caching
and selective reads are the levers, and selective reading trades
against the one-pass property of Mechanism 1.

\subsection{Trust Gradient and In-Band Governance}
\label{sec:trust}

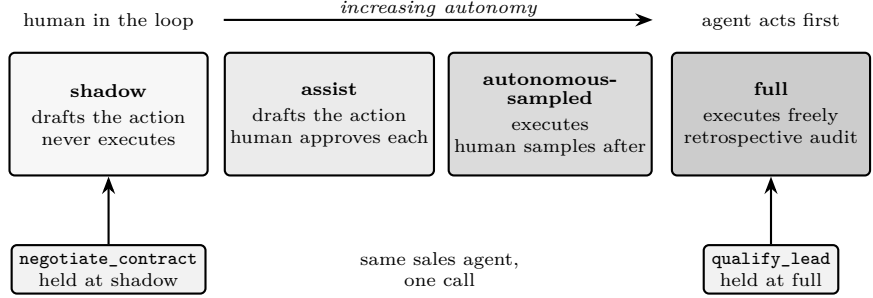
\begin{figure}[ht]
\centering
\begin{tikzpicture}[
  rung/.style={
    rectangle, draw, thick, rounded corners=2pt,
    minimum width=2.62cm, minimum height=1.62cm,
    align=center, font=\scriptsize, inner sep=2pt
  },
  skill/.style={
    rectangle, draw, thick, rounded corners=2pt, fill=black!5,
    font=\scriptsize, inner sep=3pt, align=center
  },
  flow/.style={->, thick, >=Stealth, font=\scriptsize},
  axis/.style={->, thick, >=Stealth}
]
\node[rung, fill=black!3] (shadow) at (0,0)
  {\textbf{shadow}\\[2pt] drafts the action\\never executes};
\node[rung, fill=black!8] (assist) at (2.92,0)
  {\textbf{assist}\\[2pt] drafts the action\\human approves each};
\node[rung, fill=black!14] (sampled) at (5.84,0)
  {\textbf{autonomous-}\\[-1pt]\textbf{sampled}\\[2pt]
    executes\\human samples after};
\node[rung, fill=black!20] (full) at (8.76,0)
  {\textbf{full}\\[2pt] executes freely\\retrospective audit};

\node[font=\scriptsize, anchor=south] at ($(shadow.north)+(0,0.18)$)
  {human in the loop};
\node[font=\scriptsize, anchor=south] at ($(full.north)+(0,0.18)$)
  {agent acts first};
\draw[axis] ($(assist.north west)+(0,0.42)$) to
  node[above, font=\scriptsize\itshape, inner sep=1pt]
    {increasing autonomy}
  ($(sampled.north east)+(0,0.42)$);

\node[skill] (neg) at (0,-2.05)
  {\texttt{negotiate\_contract}\\held at shadow};
\node[skill] (qual) at (8.76,-2.05)
  {\texttt{qualify\_lead}\\held at full};
\node[font=\scriptsize, align=center] (agent) at (4.38,-2.05)
  {same sales agent,\\one call};
\draw[flow] (neg) to (shadow);
\draw[flow] (qual) to (full);
\end{tikzpicture}
\caption{The per-skill trust gradient. Four rungs define what a
skill may do, from drafting with no execution (shadow) to
unsupervised execution under retrospective audit (full). Trust
attaches to skills, not to agents: one sales agent can hold
\texttt{qualify\_lead} at full and \texttt{negotiate\_contract} at
shadow on the same call. A human principal raises a level; a
deterministic controller may lower it after a severe regression,
logged as a signed commit.}
\label{fig:trust}
\end{figure}

The central governance instrument is the \emph{trust gradient}: a
per-skill trust level encoded in the skill's Markdown frontmatter.
Four levels
define what the agent may do (\cref{fig:trust}). At \textbf{shadow}
the agent drafts an action and never executes; at \textbf{assist}
a human approves each draft before execution; at
\textbf{autonomous-sampled} the agent executes and a human
spot-checks a sampled fraction; at \textbf{full} the agent
executes freely under retrospective audit.

Trust attaches to skills, not to agents. A sales agent may run
\texttt{qualify\_lead} at full autonomy while holding
\texttt{negotiate\_contract} at shadow on the same call, the pairing
shown in \cref{fig:trust}. Sensitive
actions (legal commitments, refunds above a threshold, hiring
decisions) stay gated regardless of trust elsewhere.
Trust fields accept no LLM-agent writes. A deterministic
controller may downgrade a skill after a severe regression,
logged as a Git commit; only a human principal may raise a
level. How outcomes should otherwise move a skill's level is
open. The
gradient is a substrate-encoded form of adjustable autonomy
\cite{scerri2002adjustable} and compresses classical
levels-of-automation scales \cite{parasuraman2000model}. The new
element is that the autonomy level is a versioned artifact in the
repository the agent reasons over, so every change of autonomy is
itself an auditable commit.

\subsection{Three Closed Loops, One Substrate}
\label{sec:loops}

\begin{table}[ht]
\centering
\caption{Three closed loops at three time-scales, sharing one
Markdown substrate.}
\label{tab:loops}
\footnotesize
\setlength{\tabcolsep}{4pt}
\begin{tabular}{@{}llll@{}}
\toprule
Loop   & Scale   & What closes it & Substrate artifact \\
\midrule
Action & seconds & Outcome compared to intent & Outcome appended to log file \\
Skill  & days    & Trust gradient, runtime learning
  & \begin{tabular}[t]{@{}l@{}}Playbook revised,\\ trust fields
    excepted\end{tabular} \\
Policy & weeks   & Human or governance review & Policy file updated in review \\
\bottomrule
\end{tabular}
\end{table}

Many company processes are open loops by accident: a sales reply
is sent and the outcome never examined; a refund is issued and
never audited. LLMs change the
economics of closing such loops.

The four layers map onto control roles (\cref{fig:architecture}):
the Governance Layer holds the setpoints, the Intelligence Layer is
comparator and actuator, the Knowledge Layer is the measurement
substrate, and Sync Agent ingestion is the sensor pathway. The
three loops in \cref{tab:loops} close at different time-scales but
read from and write to the same files, so one corrected outcome
serves all three: it is the next action's input, the next skill
revision's evidence, and the next policy review's pattern. This is
stigmergic coordination
\cite{theraulaz1999stigmergy,weyns2007environment}: a correction
propagates because later work reads the marks earlier
work left in the substrate. This compounding is cross-loop
coupling (Mechanism 2) at work. Coupled loops that close cheaply
and quickly can also oscillate: a skill loop may chase action-loop
noise, and a policy loop may institutionalize a transient. The
framework's dampers are the trust gradient's graduated rollout,
sampled human review, and the slower policy cadence.

\subsection{Worked Example: A Customer File}
\label{sec:example}

The following customer file is a constructed illustration, not
deployment data.

{\footnotesize
\begin{verbatim}
---
status: active
last_payment: 2026-05-01
lifetime_value_eur: 38400
plan: annual
owner: Sara Chen
---

# Acme Corp

Annual plan since March 2026; renewal on track.

## Activity log
- 2026-05-01: payment EUR 4,872 received.
- 2026-03-15: upgraded to [[Plans/Annual]].
\end{verbatim}
}

Both the sales agent and a human account owner read the same
file. The wiki links make it a node in the knowledge graph,
and the Sync Agent updates the frontmatter whenever external state
changes.

\section{Alternative Approaches}
\label{sec:alternatives}

For each of three alternative positions on agent infrastructure, we present the strongest version of the case, respond from the mechanisms of \cref{sec:position}, and state where we concede ground.

\subsection{Better Retrieval Is Sufficient}
\label{sec:alt1}

\textbf{The case.} Retrieval-augmented generation \cite{lewis2020rag} over enterprise data stores has matured rapidly: chunking, query rewriting, query expansion, learned re-rankers, and iterative retrieval close many early gaps \cite{fan2024ragsurvey}. On this view retrieval, not substrate, is the bottleneck: agents reason over existing company data while the systems of record stay untouched, and better retrieval is the cheaper, more incremental path to the same end state.

\textbf{Response.} Substrate inversion is not an argument against retrieval. Retrieval returns fragments that the agent must reassemble in working memory at every step. On H1, raising retrieval quality lifts the ceiling without removing it: better-chosen fragments are still fragments. The framework retains retrieval as a Sync Agent tool for raw records, not as the substrate. We concede that at low compositional depth the two approaches are functionally equivalent, and the choice between them is economic rather than architectural.

\subsection{Tool-Using Agents on Typed Schemas}
\label{sec:alt2}

\textbf{The case.} With function-calling interfaces \cite{schick2023toolformer}, ReAct-style reasoning loops \cite{yao2023react}, and well-typed tool contracts, agents compose typed actions into multi-step workflows without rebuilding the substrate, and the typed contract between agent and tool gives auditors a precise review surface. The multi-agent-systems community has invested decades in typed agent communication languages and shared ontologies \cite{wooldridge1995agents}. On the strongest reading, typed schemas are the solution: better tool design and training will erase the asymmetry at far lower cost than unifying everything in Markdown.

\textbf{Response.} Our position keeps typed schemas as the right tool at the action boundary. They do not, however, erase the asymmetry: the lossy case of Mechanism 1 is a decision routed field by field through a CRM, a ledger, and an email API, and better typing removes per-call ambiguity, not per-step reassembly. On H1, well-typed records are still fragments. The deeper disagreement concerns cross-loop coupling (Mechanism 2). Tool-using agents perform well within the action loop but typically distribute the skill and policy loops across the separate substrates that \cref{sec:position} described, so corrections in one loop fail to propagate to the others. The strongest variant of the objection is a hybrid that versions playbooks, policies, and outcome logs in one repository while reasoning-time reads stay tool-mediated. This captures cross-loop coupling, and the residual disagreement is then exactly Mechanism 1, the factor the benchmark in \cref{sec:call} isolates. A second variant closes the skill loop in weights by outcome-driven fine-tuning; such loops are illegible to the policy loop, reset across model versions, and instantiate the split-substrate arm of the same benchmark. We concede that at low compositional depth tool-using agents are competitive; without the hybrid, the position is predicted to become decisive at the skill-loop and policy-loop time-scales.

\subsection{Vertical AI Applications}
\label{sec:alt3}

\textbf{The case.} On this view, domain-specific applications (verticalized customer support, legal copilots, finance operations) capture value through deep specialization: each vertical builds its own optimized substrate, evaluation suite, and compliance story, and enterprise buyers prefer a finished workflow to a general substrate.

\textbf{Response.} A company running five vertical applications operates five disconnected substrates, each with its own data model, audit trail, and governance plane: the very fragmentation that cross-loop coupling predicts will impede company-level improvement. Vertical applications nonetheless fit within our framework, since they can plug into the Intelligence Layer as specialized agents that read and write the shared substrate. Single-vertical companies are well served by such applications; we concede the point. The position targets cross-functional coordination.

\section{Risks, Governance, and Scope}
\label{sec:implications}

\subsection{Risk Identification, Evaluation, and Accountability}
\label{sec:safety}

Three concerns recur: identifying risk, evaluating it, and
accounting for autonomous action. Substrate inversion makes
governance and audit structural; safety is not. Safety requires
active defense, and one risk is imported rather than removed.

\textbf{Risk identification: the substrate changes the threat model.}
The compile pass turns untrusted external content into trusted agent
context, a carrier of prompt injection \cite{perez2022}, indirect
injection in particular \cite{greshake2023}; the architecture
sharpens this threat beyond the per-call case. Persistent substrate
poisoning: a payload compiled once is re-read by every agent on
every call, so injection is durable. Cross-agent
propagation: the substrate is the message bus, so one poisoned
artifact reaches every agent that reads it. Trust-gradient
manipulation: trust fields accept no LLM-agent writes, so the
remaining attack is a poisoned compile or a bypassed commit gate,
not routine agent edits. Loop laundering: a
poisoned outcome can steer the skill loop's playbook revision.
Exfiltration: an injected directive can route data out through
authorized write channels. And the forwarding email address is
an internet-facing, unauthenticated write path into trusted context.
The Sync Agent concentrates the surface: the component compiling the
rawest input also holds every external credential. Privilege
separation is the response: an uncredentialed compile worker writes
quarantined drafts, while a credentialed deterministic action worker
executes and checks trust. The compile pass needs quarantine,
parameter validation, and directive detection, drawing on
spotlighting
\cite{hines2024spotlighting}; this compile-path screening never
finishes. Retrieval stores can also screen at ingestion; here the
screening point is single, versioned, and auditable, and a miss
persists substrate-wide. The single shared reasoning
surface co-locates control artifacts and compiled untrusted
content: policy files and untrusted summaries sit in the same
repository. CaMeL \cite{debenedetti2026camel} keeps those
channels apart; the substrate as specified does not, so it needs
trust-zone partitioning (provenance or taint labels on compiled
artifacts), an open tension.

\textbf{Risk evaluation: the trust gradient as runtime control.}
Per-skill trust levels make risk measurable at runtime; gating and
downgrade operate as in \cref{sec:trust} and \cref{fig:trust}. In-band policy provides
legibility and auditability. It cannot, by itself, enforce against
a non-compliant or injected model. Enforcement therefore
sits outside the LLM, at the action boundary: the Sync Agent's
monopoly on external calls lets deterministic code in the call path
check a skill's trust level when the action executes. Two
conditions harden the check: only designated principals may write
Governance Layer files, with one exception (a deterministic
controller may lower a skill's trust level), and the check reads
from a pinned, verified revision, not the working tree. The check gates the skill, not its
payload; egress control remains open.

\textbf{Ethical and legal accountability: audit by construction.}
Rules, approval thresholds, and trust levels are first-class
substrate artifacts; no separate rule engine drifts out of date.
Git history serves as the audit trail: every change, human or agent,
is diffable, timestamped, and revertable, and an auditor reads the
same files the agent reads. Attribution and immutability hold only
under signed commits and protected branches, controls the deployment
must enforce. Two open problems remain. Read-side access control:
the gradient gates actions, nothing yet gates reads in one shared
repository where a CRM would enforce row-level permissions;
per-directory ACLs or trust-zone sharding are candidate directions.
Erasure versus immutability: the worked example stores customer
data, and deletion under data-protection law conflicts with the
signed history the audit rests on, a conflict crypto-shredding of
per-record keys would blunt.

\subsection{Scope: Where the Position Fails}
\label{sec:scope}

The position is best suited for interpretation-heavy operations, where the
bottleneck is the agent's grasp of context and the substrate's job is
to keep that context legible. It loses for high-frequency
transactional systems (algorithmic trading, ad bidding, payments
routing), where typed-state guarantees and tight latency budgets
dominate. Nor does the position eliminate structured systems of
record: the payment processor still holds the money, and the ledger
remains authoritative.

\section{Research and Adoption Agenda}
\label{sec:call}

\textbf{Adoption path.}
Start with one department, not a company-wide cutover. Pick a
workflow that is interpretation-heavy and already has a named
owner (sales qualification, support triage, and contract review
are typical), migrate that workflow into Markdown files the
department agent reads on every call, and wire the Sync Agent
only to that department's external systems. Run the agent up the
trust gradient of \cref{sec:trust} (\cref{fig:trust}): shadow,
then assist, then autonomous-sampled, then full, with a human
principal required for every increase. The headline metric is time-to-full-autonomy,
checked against regression rate and rollback rate.

Rollback is a first-class requirement. Because the Data Layer
remains authoritative, reverting the department means resetting
the skill's trust level to shadow by a signed commit, redeploying
the previous Knowledge Layer revision, and leaving the CRM,
ledger, and mail systems untouched. The department can fall back
to its prior tools the same day. Only after the pilot holds under
a rollback drill and sampled audit should the pattern replicate
to a second department; cross-department workflows on the shared
substrate are the prize, not the starting point.

\textbf{Research agenda.}
The central item is a substrate-controlled benchmark specified
so a reader can run it. Hold the model and the task suite fixed.
Vary two factors. Representation: unstructured Markdown, a typed
schema, retrieval-augmented generation over the same documents, a
relational store, and a hybrid that versions playbooks in a
repository while reasoning-time reads stay tool-mediated. Loop
coupling: the three improvement loops of \cref{sec:position}
share one substrate, or they are split (action in agent memory,
skill in a prompt registry or fine-tune, policy in a separate
store). Every condition compiles from one source corpus, so
information content is held constant and only structure and
coupling change.

Compositional depth is the number of inter-entity relations a
task must jointly resolve (for example, a refund that depends on
the latest ticket, the payment record, and the current policy),
with source-fact count and total tokens held fixed. The
longitudinal arm runs a skill loop: a fixed agent revises the
playbook from the last $k$ outcomes, identically across
conditions, and the score is pass-rate delta per closure. The
noise floor is variance under a frozen playbook. Compounding is a
positive trend over tens of closures that clears that floor.
Replicating the design across model tiers tests whether any
plateau persists as models improve. Two further instruments are
needed: structured-merge primitives for concurrent agent edits,
and a compile-path injection benchmark that extends per-call
suites \cite{debenedetti2024agentdojo} to persistence and
cross-agent propagation.

H1 and H2 state what would falsify the position. H1 predicts that
per-step reassembly cost bounds compositional reach, so the
connected-prose condition should pull away from typed and
retrieval conditions as depth grows, with source facts held
fixed. H1 is false if the conditions stay within noise of each
other at high depth, or if typed or retrieval conditions win.
H2 predicts that under split substrates, improvement decays
toward the noise floor rather than compounding as it does on a
unified substrate, and that model upgrades alone do not rescue
the split case. H2 is false if the split condition compounds as
well as the unified one, or if a model upgrade closes the gap
without unifying the substrate. Neither outcome is yet measured;
both are the point of the benchmark.

\textbf{Open operational questions.}
Even if H1 and H2 held, a deployment would still have to answer
questions this paper does not resolve. Scalability: at what
corpus size does index-first navigation erase the one-pass
advantage of Mechanism 1, and does the per-page connectedness
claim survive? Freshness: how stale may a compiled customer file
be relative to the system of record, and who detects missed
events? Read-level authorization: the trust gradient gates
actions, not reads; what access control belongs in a shared
repository where a CRM would enforce row-level permissions?
Concurrency: the minimal write regime of \cref{sec:layers} is a
convention, not a merge semantics; what happens when two agents
edit the same playbook? Cost and latency: daily compilation is
cheap on a mid-tier model, but rereading the corpus on every
action is not; which caching and selective-read policy preserves
Mechanism 1 without blowing the seconds-scale action loop?
Recovery: if compiled files are poisoned or corrupted, what is
the restore unit (file, commit, trust zone), and how fast can
the Knowledge Layer be rebuilt from the systems of record? These
are design caveats. Treating them as solved would overclaim.

\textbf{Standards.}
For standards bodies, the open task is to specify substrate-level
audit guarantees (audit trail by construction, in-band policy,
structured trust levels) in a vendor-neutral form, so that
Git-based provenance can sit alongside log-based provenance in
compliance frameworks.

\section{Conclusion and Future Work}
\label{sec:conclusion}

Companies that deploy AI agents in sustained operation should
rebuild their cognitive substrate around representations matched
to the agent's reasoning surface, with schema translation confined
to the action boundary. Markdown is the instantiation available
today, not a proven primitive, and superiority over typed-tool,
retrieval, or hybrid stacks remains a hypothesis, not a result.
The commitment concerns the reasoning surface, not the framework,
runtime, or model class.
Its operational form is the four-layer framework, the Sync Agent,
the per-skill trust gradient, and the
three closed loops. Policy legibility and a single audit surface
then follow by construction. Safety and attribution additionally
require enforcement at the action boundary, signed commits, and a
guarded compile path (\cref{sec:safety}).

Five questions remain open. \textbf{(1) Threat model.} Compile-path
injection demands a standing threat-model review and a dedicated
benchmark.
\textbf{(2) Cost envelope.} Continuous compilation and continuous
reasoning are LLM-heavy; cost levers such as model routing and
incremental compilation need calibration on real workloads.
\textbf{(3) Control and concurrency.} Agent activation and the
orchestrator's policy remain unspecified, and edits beyond the
minimal regime of \cref{sec:layers} exceed default Git merge
semantics. \textbf{(4) Skill
discovery at the long tail.} Novel goals need a compose-or-escalate
path that the framework leaves under-specified. \textbf{(5)
Empirical validation.} The position has not yet been benchmarked
against the alternatives of \cref{sec:alternatives}, and the
experiment that would test hypotheses H1 and H2 is specified in
\cref{sec:call}. None of these are blockers for stating the
position; they are the work of testing it.

We offer the substrate-inversion position as work in progress:
something to argue with and to test in deployment.

\bibliographystyle{splncs04}
\bibliography{references}

\end{document}